\def\a{\alpha}\def\d{\delta}
\def\f{\phi}\def\h{\theta}
\def\k{\kappa}\def\m{\mu}\def\n{\nu}\def\o{\omega}\def\q{\psi}
\def\y{\eta}

\def\D{\Delta}\def\F{\Phi}\def\G{\Gamma}\def\L{\Lambda}

\def\V{\varphi}

\def\de{\partial}
\def\inf{\infty}\def\id{\equiv}\def\ha{{1\over 2}}

\def\({\left(}\def\){\right)}\def\[{\left[}\def\]{\right]}

\def\mn{{\mu\nu}}\def\ij{{ij}}

\def\tran{transformations }\def\coo{coordinates }

\def\rep{representation }

\def\pb{Poisson brackets }

\def\poi{Poincar\'e }

\def\trl{transformation law }

\def\cor{commutation relations }

\def\kp{$\k$-\poi }\def\km{$\k$-Minkowski }

\def\section#1{\bigskip\noindent{\bf#1}\smallskip}

\def\subsection#1{\smallskip\noindent{\it#1}\smallskip}

\def\PL#1{Phys.\ Lett.\ {\bf#1}}
\def\PRL#1{Phys.\ Rev.\ Lett.\ {\bf#1}}
\def\PR#1{Phys.\ Rev.\ {\bf#1}}\def\CQG#1{Class.\ Quantum Grav.\ {\bf#1}}

\def\JoP#1{J.\ Phys.\ {\bf#1}} \def\IJMP#1{Int.\ J. Mod.\ Phys.\ {\bf #1}}

\def\AoP#1{Ann.\ Phys.\ {\bf#1}}

\def\JHEP#1{JHEP\ {\bf#1}}\def\EPJ#1{Eur.\ Phys.\ J.\ {\bf#1}}

\def\hep#1{{\tt hep-th/#1}}\def\arx#1{{\tt arXiv:#1}}

\def\EPL#1{EPL\ {\bf#1}}

\def\ref#1{\medskip\everypar={\hangindent 2\parindent}#1}
\def\beginref{\begingroup
\bigskip
\centerline{\bf References}
\nobreak\noindent}
\def\endref{\par\endgroup}

 \font\ttl=cmtt10 scaled 1200
\def\msm{\(1-{p_0\over\k}\)}\def\cP{{\cal P}}\def\cX{{\cal X}}
\def\po{{p_0\over\k}}\def\ar{{\a\over r}}\def\Eo{{E\over\k}}\def\ij{{ij}}
\def\tx{\hat x}\def\tp{\hat p}\def\PI{{\ttl P}}\def\TI{{\ttl T}}\def\CI{{\ttl C}}
\def\hp{\tilde p}

{\nopagenumbers
\line{}
\vskip30pt
\centerline{\bf CPT breaking in noncommutative Magueijo-Smolin model}

\vskip60pt
\centerline{
{\bf S. Mignemi}\footnote{$^\ddagger$}{e-mail: smignemi@unica.it}}

\vskip10pt

\smallskip
\centerline{$^1$Dipartimento di Matematica e Informatica, Universit\`a di Cagliari}
\centerline{via Ospedale 72, 09124 Cagliari, Italy}
\smallskip
\centerline{$^2$INFN, Sezione di Cagliari, Cittadella Universitaria, 09042 Monserrato, Italy}

\vskip80pt

\centerline{\bf Abstract}
\medskip
{\noindent We review an instance of noncommutative geometry based on a specific realization
of the model of doubly special relativity proposed by Magueijo and Smolin (MS) on
noncommutative spacetime. In particular, we discuss
the Hopf algebra associated to it, which has not been considered in the literature till now.
We show that the momentum sector of this model can be viewed as a particular basis of the \kp model.
An interesting property is that the MS Hamiltonian is not invariant under the reversal of
the sign of the energy, and in particular it is not invariant under the standard definition of charge conjugation.
Therefore, if following Dirac one identifies the negative energy states with antiparticles, their mass differs
from that of particles.
We examine the possible consequences of this fact in the context of first quantization
and discuss its interpretation from the point of view of quantum field theory, taking into account possible
alternative definitions of charge conjugation proposed in the noncommutative framework.
}
\vskip10pt
{\noindent

}
\vfill\eject\

}

\section{1. Introduction}

Twenty years ago Magueijo and Smolin (MS) [1] proposed one of the first models of what became known as
doubly special relativity (DSR) [2].
DSR theories postulate the existence of a fundamental energy scale, originating from quantum gravity, whose invariance,
together with the requirement of the validity of  the principle of relativity, implies a nonlinear deformation of
the action of the Lorentz group on momentum space at extremely high energies.
In particular, the deformation proposed in [1] leaves invariant the Planck energy $\k=\sqrt{\hbar c^5/G}$.

As well known, DSR mainly deals with momentum space, while the nature of spacetime is usually not specified by the theory.
Thus in the original papers on the MS model [1] the structure of spacetime was not discussed, even if it was implicitly
assumed to be commutative.

Later, some proposals were advanced for a realization of the model on noncommutative spacetime [3-7].
This looks natural since DSR models are usually associated with noncommutative spacetimes, that also entail a fundamental
scale.
Among these representations, the most interesting was in our opinion the one proposed by Granik [3] (see also [5,7]), because
of its simplicity. The key request for obtaining this realization is that the kinematical definition of the
3-velocity (phase velocity)\footnote{$^1$}{We adopt the following notation: $i,j,\dots=1,2,3$; $\m,\n,\dots=0,1,2,3$, metric signature
$(-1,1,1,1)$ and natural units $c=\hbar=1$.}, $v^i={\dot x^i\over\dot x^0}$, when written in terms of the momenta, give rise to the
same result as special relativity, $v_i={p_i\over p_0}$,  and consequently that the speed of particles does not depend on
their energy\footnote{$^2$}{The issue of the definition of velocity in DSR has been discussed in several papers [8].}.
This can be achieved only if the position coordinates are noncommutative.

More recently, the Granik realization was included in a class of models obtained from an
interpolation of Jordanian twists [9-10], and then was studied more explicitly in [11-12].
Actually, the \cor of spacetime \coo proposed by Granik are the same as those of \km spacetime. Hence, from the viewpoint of
noncommutative geometry the phase space of the Granik realization is simply a particular basis of the Hopf algebra of the \kp model [13].
The explicit form of the coalgebra in this basis has however not been studied till recently, when the twist formalism was used [11-12].
However, it must be noted that in the Granik basis the Lorentz sector of the MS model cannot be obtained from the \kp algebra through
the same transformations as the momentum sector. This fact will give rise to serious problems when dealing with discrete symmetries.

In fact, a troublesome property of the MS model is the lack of invariance of its Casimir operator
$$C={-p_0^2+p_i^2\over(1-p_0/\k)^2},\eqno(1)$$
for $p_0\to-p_0$ [14]\footnote {$^3$}{As recently observed in [15], this property is common to other noncommutative models.}.
The Casimir operator is usually identified with minus the mass square $-m^2$ of the relativistic particle and gives rise
to the Hamiltonian constraint and to the dispersion relation for particles.

By definition, $C$ is invariant under deformed Lorentz transformations, and is also invariant under the standard
definition of the discrete symmetries \PI\ and \TI, which do not affect $p_0$, but the Klein-Gordon equation is not invariant under
complex conjugation and hence under the standard definition of charge conjugation. Moreover, writing $p_0$ in terms of $p_i$ and the
mass $m$ of a particle, one gets [14]
$$p_0=-{m^2/\k\pm\sqrt{p_i^2(1-m^2/\k^2)+m^2}\over1-m^2/\k^2}.\eqno(2)$$
Hence the relativistic energy $p_0$ can take two different values for a given 3-momentum. Adopting the Dirac interpretation of negative
energy states of the first quantized  theory, one would associate the two different energies to particles and antiparticles, thus breaking
the charge conjugation symmetry, as first noticed in the nonrelativistic limit in [14].
In the commutative case there seems to be no way out of this conclusion, but even adopting different interpretations it is not clear how to
single out the correct sign in (2).

However, a rigorous discussion of this topic must be afforded in the context of a quantum field theory (QFT), which in a noncommutative
setting is usually implemented by means of the formalism of Hopf algebras [16].
In particular, in [17] it has been argued that in such formalism the discrete symmetries act in a nontrivial way, that involves the
antipodes of the Hopf algebra, and this idea has been subsequently implemented in [18,19].
In any case, the standard \CI\PI\TI\ theorem is not necessarily valid in noncommutative field theory [19].

In this paper, we review the MS model in the Granik realization and discuss its Hopf algebra, investigating its behavior under discrete
symmetries.
We also consider the possible quantum mechanical \rep of the Heisenbeg algebra of the model on a Hilbert space, and use it to study
some elementary first-quantized models.
Finally, in the last sections we discuss QFT in the Hopf algebra approach and considering the alternative definition of charge
conjugation based on the antipode.

Our conclusion is that the Granik realization cannot satisfy charge conjugation invariance because, in order to get the postulated
commutation relations starting from the bicrossproduct basis, the momentum and the Lorentz sectors are obtained by applying different
transformations.
However, different realizations are available that, assuming an alternative deformation of the Heisenberg algebra, avoid this problem [4]
and allow to restore a formal invariance by adopting a nonstandard definition of charge conjugation.
One could therefore consider these realizations as more natural from an algebraic point of view, although their form is rather involved.

However, this does not solve completely the problem of charge conjugation, since the physical interpretation of the 4-momentum
of an antiparticle in terms of the antipode is not obvious.
Related to this is the definition of a nonrelativistic limit of the theory. In fact, it has been recently observed that even in the
bicrossproduct basis the dispersion relation  are not invariant under $p_0\to-p_0$, leading to similar problems [17].

\section{2. The MS model}
DSR models can be characterized through the deformation of the action of the Lorentz group on the momenta.
In particular, the infinitesimal deformations can be described by means of
the commutators between the Lorentz algebra generators $J_\mn=x_\m p_\n-x_\n p_\m$ and the momentum generators $p_\m$.
In the MS model, while the action of rotations $J_\ij$ is undeformed, that of boosts $J_{i0}$ is different from the case of special
relativity and is given by [1]
$$\d_ip_0=[J_{i0},p_0]=i\msm p_i,\qquad\d_ip_j=[J_{i0},p_j]=i\(\y_\ij p_0-{p_ip_j\over\k}\),\eqno(3)$$
reducing to the undeformed action only for small 4-momentum.
Because of (3), the \poi algebra is deformed and its Casimir invariant $C$ is given by (1).
The Casimir operator diverges for $p_0\to\k$, and therefore $\k$ must be interpreted as an upper limit for the energy of an elementary
particle. Moreover, as noted before, $C$ is not invariant for $p_0\to-p_0$.

The transformations (3) do not fix the action of the \poi algebra on the spacetime \coo nor the deformation of the Heisenberg algebra,
so that one can consider different realizations of the model on spacetime [3-7], including a trivial commutative one [1].

In this paper, we consider the proposal of Granik [3], that seems to be the most natural.
According to it, the Heisenberg algebra is deformed as
$$[x_i,x_0]=i{x_i\over\k},\quad[x_0,p_i]=i{p_i\over\k},\quad[x_i,p_j]=i\d_{ij},\quad[x_0,p_0]=-i\(1-{p_0\over\k}\),\eqno(4)$$
so that the position operators satisfy the same \cor as in the \km spacetime. Notice however that infinite deformations of the
Heisenberg algebra are compatible with the postulated \cor for position operators.

Given (4) and the Jacobi identities, one can deduce the infinitesimal transformations of the coordinates under boosts [5]
$$\d x_0=[J_{i0},x_0]=-i\(x_i-{p_i\over\k}x_0\),\qquad\d x_i=[J_{i0},x_j]=-i\(\y_\ij x_0-{p_i\over\k}x_j\).\eqno(5)$$

The classical dynamics of a free relativistic particle can be obtained using the deformed \pb corresponding to the commutators (3)
and writing the Hamiltonian constraint as $C+m^2=0 $,
with $C$ the Casimir operator (1). However, a shortcut to obtain the correct equations of motion is to  use
the deformed \pb corresponding to the commutators (3), with Hamiltonian  $H={-C\over2m}$.
The Hamilton equations then read
$$\dot x_\m={p_\m\over m\msm^2},\qquad\dot p_\m=0.\eqno(6)$$
and hence $v_i\id{\dot x^i\over\dot x^0}={p_i\over p_0}$. Note however that the group velocity for massive particles is deformed,
$$v^G_i\id{\de p_0\over\de p_i}={p_i\over\sqrt{m^2+\(1-{m^2\over\k^2}\)p_i^2}}={p_i\over\(1+{m^2\over\k^2}\)p_0-{m^2\over\k}},\eqno(7)$$
while for massless particles it coincides with the phase velocity.

From the \trl (3) and (5) one can also prove the existence of an invariant "metric" in phase space [5]
$$ds^2=\msm^2(-dx_0^2+dx_i^2)={-p_\m p^\m\over m^2}(-dx_0^2+dx_i^2).\eqno(8)$$

\bigbreak
\section{3. Quantum mechanics of the MS model}
In this section we discuss the MS model in the context of quantum mechanics. To this purpose, it is useful
to find a quantum mechanical \rep of the  Granik algebra in a Hilbert space [7,10].
The \cor of the Granik algebra (4) can be obtained in infinitely many ways by nonlinear transformations of
canonical phase space coordinates $\tx_\m$, $\tp_\m$ obeying
$$[\tx_\m,\tx_\n]=[\tp_\m,\tp_\n]=0,\qquad[\tx_\m,\tp_\n]=i\,\y_\mn.\eqno(9)$$
Two main cases have been studied in the literature:
$$p_\m=\tp_\m,\qquad x_0=\tx_0-{1\over\k}\tx^\m\tp_\m,\qquad x_i=\tx_i,\eqno(10)$$
introduced in [10], and
$$p_\m={\tp_\m\over1+\tp_0/\k},\qquad x_\m=\(1+{\tp_0\over\k}\)\,\tx_\m,\eqno(11)$$
proposed in [7].

The Klein-Gordon (KG) equation has the general form [14]
$$\left[-p_0^2+p_i^2+m^2\left(1-\po\right)^2\right]\f=0,\eqno(12)$$
and can be written in a Hilbert space using the previous realizations, either in a momentum or in a position representation.

For the realization (10), introducing a momentum variable $\cP_\m$, one has
$$p_\m=\cP_\m,\qquad x_0=i\,{\de\over\de\cP_0}-{i\over\k}\cP_\m\,{\de\over\de\cP_\m},\qquad x_i=i\,{\de\over\de\cP_i},\eqno(13)$$
and Hilbert space measure $d\m=d^4\cP$.
The KG equation takes the form $(\cP_0^2-\cP_i^2)\q=m^2(1-\cP_0/\k)^2\q$, with solution
$$\q\propto\d\(\cP_0^2-\cP_i^2-m^2\(1-{\cP_0\over\k}\)^2\).\eqno(14)$$

\bigskip
The realization (11) in momentum space is given instead by
$$p_\m={\cP_\m\over1+\cP_0/\k},\qquad x_\m=i(1+\cP_0/\k)\,{\de\over\de\cP_\m},\eqno(15)$$
where $-\inf<\cP_i<\inf$, $0<\cP_0<\inf$.
In this case, the measure for which the operators $x_\m$ are symmetric is given by
$$d\m={d^4\cP\over1+\cP_0/\k},\eqno(16)$$
The KG equation reads $(\cP_0^2-\cP_i^2)\q=m^2\q$ and has the obvious solution
$$\q\propto\d(\cP_0^2-\cP_i^2-m^2).\eqno(17)$$

Analogously, one may write the realizations in position space. The most useful realization is given by eq.~(10).
In terms of position variables $\cX_\m$, one has
$$p_\m=-i{\de\over\de\cX^\m},\qquad x^0=\cX^0+{i\over\k}\cX^\m{\de\over\de\cX^\m},\qquad x^i=\cX^i.\eqno(18)$$
and hence the KG equation reads
$$\left[{\de^2\over\de\cX_0^2}-{\de^2\over\de\cX_i^2}+m^2\left(1+{i\over\k}\,{\de\over\de\cX_0}\right)^2\right]\f=0.\eqno(19)$$

Some nontrivial consequences follow from the lack of invariance of the Hamiltonian for $p_0\to-p_0$.
The KG equation is invariant under the standard definitions of parity, $x_0\to x_0$, $x_i\to-x_i$, implying \PI$(p_0)=p_0$, \PI$(p_i)=-p_i$,
and time reversal, $x_0\to-x_0$,  $x_i\to x_i$ together with complex conjugation, implying \TI$(p_0)=p_0$, \TI$(p_i)=-p_i$,
but it is not invariant under the standard definition of charge conjugation. In fact, the complex conjugate of (19) gives
$$\left[{\de^2\over\de\cX_0^2}-{\de^2\over\de\cX_i^2}+m^2\left(1-{i\over\k}\,{\de\over\de\cX_0}\right)^2\right]\f^\dagger=0.\eqno(20)$$
and therefore $\f^\dagger$ does not satisfy the same equation (19) as $\f$.

One can see this also by explicitly solving the equation. Its plane wave solutions are given by
$$\q=e^{-i(\o^\pm\cX^0-k_i\cX^i)},\eqno(21)$$
where
$$\o^\pm=-{m^2/\k\pm\sqrt{k^2(1-m^2/\k^2)+m^2}\over1-m^2/\k^2}=\pm\sqrt{m^2+k^2}-{m^2\over\k}+o\({1\over\k^2}\).\eqno(22)$$
The two values of $\o^\pm$ correspond to positive and negative energy states and following Dirac might be interpreted as belonging to particles
and antiparticles.

The splitting of the absolute value of the energy was originally observed in the nonrelativistic limit in ref.~[14].
The origin of this fact is that the invariance of the Klein-Gordon equation for $p_0\to-p_0$ is broken.
In particular, it follows that the rest mass $m^-_0$ of antiparticles  differs from that of particles, $m^+_0$, namely,
$$m^\pm_0={m\over1\mp{m\over\k}}.\eqno(23)$$

Of course, a more rigorous treatment of this property should be undertaken in the context of QFT.
Also, the definition of the conjugated field in presence of $\k$-deformation could be nontrivial [17].
We shall address these problems in the following section.
\medskip
For the moment, we attempt to
investigate charge conjugation in more depth, considering the hydrogen atom, in the the \rep (18). Since the theories studied in this
paper are intrinsically relativistic, we consider its relativistic version in a simplified setting, where the fermionic nature of the electron is
neglected [20].
Therefore, we analyze the first-quantized  Klein-Gordon equation for a particle of Casimir mass $m$ and unit charge in a central electric field.

If we introduce a minimal coupling with the Maxwell field, $p_\m\to p_\m-eA_\m$, and consider a central electric field of the form
$A_0={e\over r}$, the KG equation becomes
$$\left[-\left(p_0-\ar\right)^2+p_i^2+m^2\left(1-{1\over\k}\(p_0-{\a\over r}\)\right)^2\right]\f=0,\eqno(24)$$
where $\a$ is proportional to the central electric charge and $r=\sqrt{\cX_i^2}$. A similar equation was studied in [12]. We now make the substitutions (18)
and pass to spherical coordinates, imposing the ansatz
$$\f(\cX_0,\cX_i)=\sum_{ml}e^{-iE\cX_0}Y_{lm}(\h,\V)\psi_{lm}(r),\eqno(25)$$
where $Y_{lm}(\h,\V)$ are spherical harmonics. We obtain from (24)
$$\[{\de^2\over\de r^2}+{2\over r}{\de\over\de r}+{\a^2\(1-{m^2\over\k^2}\)-l(l+1)\over r^2}-{2\a\(E\(1-{m^2\over\k^2}\)+{m^2\over\k}\)\over r}+E^2-m^2\left(1-\Eo\right)^2\]\q_{lm}=0.\eqno(26)$$

Up to a normalization constant, the regular solutions of this equation can be written as
$$\q_{lm}=e^{-\D r}r^{\L-1\over2}\,L^\L_n(2\D r),\eqno(27)$$
where
$$\D=\sqrt{m^2\(1-\Eo\)^2-E^2},\qquad\L=2\sqrt{\(l+\ha\)^2-\a^2\(1-{m^2\over\k^2}\)},\eqno(28)$$
and $L^\a_n$ are generalized Laguerre  polynomials. Regularity is achieved if $n$ is an integer given by
$$n=-{1+\L\over2}-{\a E\(1-{m^2\over\k^2}\)+{m^2\over\k}\over\D}.\eqno(29)$$

Solving (29) for $E$, one can obtain the exact spectrum of the Klein-Gordon equation (24),
$$E_{nl}={1\over1-{m^2\over\k^2}}\[\pm {m\over\sqrt{1+{\a^2\over N^2}\(1-{m^2\over\k^2}\)}}+{m^2\over\k}\]=m\[\pm\(1-{\a^2\over2N^2}\)+{m\over\k}\]+o\({1\over\k^2}\),\eqno(30)$$
with
$$N=n+\ha+\sqrt{\(l+\ha\)^2-\a^2\(1-{m^2\over\k^2}\)}.\eqno(31)$$
 The leading corrections to the commutative relativistic spectrum are of order $m/\k\sim10^{-23}$.
As for the free KG equation, one  can observe a breaking of the particle/antiparticle symmetry, but even the spectrum of the positive energy
states is deformed.

\section{4. Hopf algebra}
In the following section, we shall investigate the Granik basis of MS model in the context of QFT. We shall do this employing the formalism of Hopf algebras [16], which is
at the basis of the interpretation of \kp noncommutative field theory [21-23].

As mentioned above, the \cor between spacetime \coo are the same as in \km spacetime.
The momentum sector of the MS model can therefore be obtained through a nonlinear transformation of the parametrization of other \kp bases.
In particular, the Granik basis is related by a change of variables to the bicrossproduct basis [24]. Denoting by $P_\m$
the momentum in bicrossproduct basis, its  commutation relations with \km position operators are
$$[x_0,P_i]=i{P_i\over\k},\quad[x_i,P_j]=i\d_{ij},\quad[x_0,P_0]=-i,\eqno(32)$$
and it is easy to find the relation
$$p_0=\k\(1-e^{-P_0/\k}\),\qquad p_i=P_i.\eqno(33)$$

This permits to construct the Hopf algebra associated to the Granik basis.
In fact, in the bicrossproduct basis the coproduct is
$$\D P_0=P_0\otimes1+1\otimes P_0,\qquad \D P_i=P_i\otimes 1+e^{-P_0/\k}\otimes P_i,\eqno(34)$$
and the antipode
$$S(P_0)=-P_0,\quad S(P_i)=-P_ie^{P_0/\k}.\eqno(35)$$

Using the homeomorphism property of Hopf algebras, one can then immediately obtain the
coproduct of momenta in the Granik basis,
$$\D p_\m=p_\m\otimes1+\msm\otimes p_\m,\eqno(36)$$
and the antipode
$$S(p_\m)=-{p_\m\over1-p_0/\k}.\eqno(37)$$
Clearly, $S^2(p_\m)=p_\m$.
Moreover, from (36) follows the addition law of momenta
$$p_\m^{(1)}\oplus p_\m^{(2)}=p_\m^{(1)}+p_\m^{(2)}-{p_0^{(1)}p_\m^{(2)}\over\k},\eqno(38)$$
and the star product
$$e^{ik\cdot x}\star e^{iq\cdot x}=e^{i\big(k+\(1-{k_0\over\k}\)q\big)\cdot x}.\eqno(39)$$

However, the extension of the relations (33) to the Lorentz sector does not give rise to the \tran (3).
To get the correct Lorentz transformations associated to the MS basis, one must also map the bicrossproduct
boost generators $M_{0i}$ to new ones given by $M_{0i}\to J_{0i}=M_{0i}-x^\m p_\m\ p_i$, while the rotation generators
$M_\ij$ stay unaltered.  A similar problem, related to the way in which
the MS model is defined [1] also appears in the approach of  [11] and [12], based on twist.

A consequence of this fact is that, contrary to what happens in the bicrossproduct basis,  the antipode (37) does not
leave the Casimir operator (1) invariant.
In fact,
$$S\[-p_0^2+p_i^2\over\msm^2\]=-p_0^2+p_i^2,\eqno(40)$$
while an expression invariant  under the antipodal transformation would be
$$\hat C={-p_0^2+p_i^2\over1-p_0/\k}.\eqno(41)$$

This fact is important because recently it has been argued that the correct definition of discrete symmetries
in the context of the Hopf algebra formalism must be given in terms of the antipode as\  \PI$p_0=-S(p_0)$, \PI$p_i=S(p_i)$,
\TI$p_0=-S(p_0)$, \TI$p_i=S(p_i)$ and similarly for charge conjugation [17]. Clearly, with this interpretation, also
\PI\ and \TI\ would be broken in the KG equation (12).

\bigbreak
\section{5. Quantum field theory}
In order to define a QFT for MS spacetime, we follow [22] (see also [21,23]).
We adopt the time to the right ordering $:e^{ip_\m x_\m}:=e^{ip_ix_i}e^{-ip_0x_0}$.
With this definition,
$$:e^{ip^{(1)}_\m x_\m}:\ :e^{ip^{(2)}_\m x_\m}:\ =\ :e^{i(p^{(1)}_\m\oplus p^{(2)}_\m)x_\m}:.\eqno(42)$$
We define the Fourier transform as
$$\F(x)=\int d\m(p)e^{ip_ix_i}e^{-ip_0x_0}\,\tilde\F(p),\eqno(43)$$
where the measure
$$d\m(p)={d^4p\over(1-p_0/\k)^5}\eqno(44)$$
is the natural measure for momentum space, invariant under deformed Lorentz \tran (3).\footnote{$^4$}{The same measure can also be obtained by
the methods of [25] for the definition of Hermitian realizations of Minkowski spacetime.}

The inverse Fourier transform is then
$$\tilde\F(p)=\int d^4xe^{iS(p_i)x_i}e^{-iS(p_0)x_0}\,\F(x)=\int d^4xe^{-i{p_ix_i\over1-p_0/\k}}e^{i{p_0x_0\over1-p_0/\k}}\,\F(x),\eqno(45)$$
where we have used the relation
$$e^{ip_ix_i}e^{-ip_0x_0}=e^{iS(p_0)x_0}e^{-iS(p_i)x_i}=e^{-{ip_0x_0\over1-p_0/\k}}e^{ip_ix_i\over1-p_0/\k}.\eqno(46)$$

The Hermitian conjugate field is defined as
$$\eqalign{\F^\dagger(x)&\equiv\int d\m(p)e^{ip_0x_0}e^{-ip_ix_i}\,\tilde\F^*(p)=\int{d^4p\over(1-p_0/\k)^5}\,e^{{ip_ix_i\over1-p_0/\k}}e^{-ip_0x_0\over1-p_0/\k}\,\tilde\F^*(p)\cr
&=\int d^4p\,e^{-{ip_ix_i}}e^{ip_0x_0}\,\tilde\F^*(S(p)).}\eqno(47)$$

One gets therefore
$$\tilde\F^\dagger(p)=\msm^5\tilde\F^\star(S(p)),\eqno(48)$$
and real fields are defined by
$$\tilde\F(p)=\msm^5\tilde\F^\star(S(p)).\eqno(49)$$
It is evident that complex conjugate fields are defined in a highly nontrivial way in momentum space.


The action for a free quantum scalar field of mass $m$ invariant under deformed Lorentz transformations
 can be written in momentum space as [22,23]
$$\ha\int d\m\,\tilde\F^\dagger(p)(C+m^2)\tilde\F(p)=\ha\int d^4p\,\tilde\F^*(S(p))(C+m^2)\tilde\F(p)\eqno(50)$$
where $C$ is the Casimir operator (1). While the scalar product
$$\int d\m\,\tilde\F^\dagger(p)\, \tilde\F(p)\eqno(51)$$
is by definition invariant under Hermitian conjugation, the Casimir operator is not, because of (40).
Hence the invariance under charge conjugation is broken in the Granik realization.
On the other hand, the expression (41) which is invariant under charge conjugation
is not invariant under deformed \poi transformations.

For what concerns parity and time reversal, instead, the standard definition of these symmetries,
leaves the action (49) invariant.
As noticed before, however, the definition of the discrete symmetries in \kp models is not well
established. For example, in ref.~[17] it is argued that the discrete \tran must be defined in terms
of the antipode. The QFT of a scalar field has been studied in [18,19] from this point of view.
It is difficult to compare the formalism used in that paper with ours, because several nontrivial
assumptions and interpretations of the formalism are adopted there, for example different transformation
properties for particles and antiparticles are assumed (which anyway looks like an implicit
breaking of invariance).
In any case, it appears that with the definitions of [17], also the \PI\ and \TI\ symmetries
are broken, since the Lorentz-invariant action (49) does not respect them, and
the behavior under charge conjugation is even worst in that case.


\section{6. Kowalski-Glikman-Nowak basis}
In sect.~4 we have observed that under the map (33) the the Lorentz sector of the Granik basis is not isomorphic
to that of the bicrossproduct basis. There is however a basis, introduced in [4], in which the \cor are slightly different
from the ones of the Granik basis and give the correct MS Lorenz \tran when a suitable isomorphism is
applied to the bicrossproduct basis. This is obtained by setting
$$\tilde p_0={\k\over2}\(1-e^{-2P_0/\k}+{P_i^2\over\k^2}\),\qquad\tilde p_i=P_i\eqno(52)$$
the \cor between the positions and momenta read then
$$[x_0,\hp_i]=i{\hp_i\over\k},\quad[x_i,\hp_j]=i\d_{ij},\quad[x_0,\hp_0]=-i\(1-{2\hp_0\over\k}\).\eqno(53)$$
(Note the factor 2 in the last relation).
The coproduct and the antipode are rather cumbersome,
$$\D\hp_0=\hp_0\otimes1+\G^2\otimes\hp_0+{\G\over\k}\hp_k\otimes\hp_k,\qquad \D\hp_i=\hp_i\otimes 1+\G\otimes\hp_i,\eqno(54)$$
$$S(\hp_0)=-{1\over\G}\(\hp_0-{\hp_i^2\over\k}\)\qquad S(\hp_i)=-{\hp_i\over\sqrt\G},\eqno(55)$$
where
$$\G=1-{2\hp_0\over\k}+{\hp_i^2\over\k^2}.\eqno(56)$$
The Lorentz \tran are still assumed to take the form (3), so that the Casimir operator $C$ is again (1), but now it is invariant
also under the antipodal transformations (55) and therefore the dispersion relation is the same for particles and antiparticles
and the action (50) is formally invariant under \CI\PI\TI\ \tran defined in terms of the antipode.
Notice however that the expression of the complex conjugated field, obtained with the same procedure of the previous
section, is in this case very awkward, and we do not report it here.

\section{7. Discussion}
The MS model [1] was one of the first proposals of DSR [2],
a theory which introduces the Planck mass as a fundamental constant giving rise to a deformation of the dispersion
relations of particles.
Its most natural realization is in the context of a noncommutative spacetime [3].
Some time ago it was shown that in the nonrelativistic limit of the MS model particles and antiparticles
possess a different rest mass, indicating a possible breaking of \CI\PI\TI\ invariance [14].

In this paper, we have discussed the relativistic quantum mechanics of the
MS model from the point of view of noncommutative geometry in the Granik basis [3],
exploiting its isomorphism with the bicrossproduct basis of \kp.
From our results it appears that the charge conjugation invariance, and consequently \CI\PI\TI, is broken,
as already was implicit from the nonrelativistic limit of the theory [14].
While this is evident when one uses the standard definitions of discrete symmetries, it is not so obvious if one adopts the
alternative definition proposed some time ago for \km spacetime, given in terms of the antipode, that restores the
invariance in the \km case, at least in a formal way [17].
However, also with this definition, the invariance in the Granik basis is broken. This can be ascribed to the fact that the
momentum and Lorentz sector of the Granik basis are related to the bicrossproduct basis by different transformations.

However, using the alternative realization of the MS model proposed in [4], it is possible to obtain a formal
invariance under \CI\ if the Hermitian conjugation is defined through the antipode of the Hopf algebra, at the cost
of having an awkward expression for the complex conjugated fields.
The physical interpretation of this result is however not clear, in particular in what concerns the behaviour of
antiparticles in the nonrelativistic limit: what is the meaning of the different expressions of the energy and momentum
of particles and antiparticles? Which is the correct sign in (2)?

As was recently noticed, an analogous situation holds in the nonrelativistic limit of other models of DSR [15].
It is therefore of fundamental importance to understand how to get a nonrelativistic limit compatible with a
definition of discrete symmetries in terms of antipode.
The solution of such problem should be independent of the peculiar case investigated in the present paper.
In this perspective, it could be interesting to investigate \CI\PI\TI\ using alternative definitions of
noncommutative QFT as the ones proposed in [18] or [19], and we plan to study this point in future.

Of course the corrections predicted by the theory are sensible only at Planck scales, but experiments with
elementary particles are not far from the possibility of testing the presumed mass differences between
particles and antiparticles [14].

Finally, we observe that different attitudes can be taken about the interpretation of our results:
one may dismiss Granik model as unphysical, but one may also speculate that the breaking of the \CI\ invariance
in models of the evolution of the early universe could lead to an explanation of the observed particle/antiparticle
asymmetries. Naturally, the last word is left to observations.

\section{Ackowledgement}
I wish to thank Giacomo Rosati and Andjelo Samsarov for illuminating discussions.
The author would like to acknowledge the contribution of GNFM and of COST Action CA23130.
\beginref
\ref [1] J. Magueijo and L. Smolin, \PRL{88}, 190403 (2002); \PR{D67}, 044017 (2002).
\ref [2] G. Amelino-Camelia, \PL{B510}, 255 (2001); \IJMP{D11}, 35 (2002).
\ref [3] A. Granik, \hep{0207113} (2002).
\ref [4] J. Kowalski-Glikman and S. Nowak, \IJMP{D12}, 299 (2003); \CQG{20}, 4799 (2003).
\ref [5] S. Mignemi, \PR{D68}, 065029 (2003).
\ref [6] D. Kimberly, J. Magueijo and J. Medeiros, \PR{D70}, 084007 (2003).
\ref [7] S. Ghosh and P. Pal, \PR{D75}, 105021 (2007).
\ref [8]  P. Kosi\'nski and P. Ma\'slanska, \PR{D68}, 067702 (2003).
S. Mignemi, \PL{A316}, 173 (2003).
M. Daszkiewicz, K. Imilkowska and J. Kowalski-Glikman, \PL{A323}, 345 (2004).
\ref [9] S. Meljanac, M. Stojic, \EPJ{C47}, 531 (2006);
S. Kresic Juric, S. Meljanac, M. Stojic, \EPJ{C51}, 229 (2007); 
A. Borowiec and A. Pachol, \PR{D79}, 045012 (2009).
\ref [10] S. Meljanac, D. Meljanac, A. Pachol and D. Pikuti\'c, \JoP{A50}, 265201 (2017).
\ref [11] P. Aschieri, A. Borowiecz and A. Pachol, \JHEP{10}, 152 (2017).
\ref [12] S. Meljanac, D. Meljanac, S. Mignemi and R. \v Strajn, \PR{D99}, 126012 (2019).
\ref [13] J. Lukierski, A. Nowicki, H. Ruegg and V.N. Tolstoy, \PL{B264}, 331 (1991);
J. Lukierski, A. Nowicki and H. Ruegg, \PL{B293}, 344 (1992).
\ref [14] M. Coraddu and S. Mignemi, \EPL{91}, 51002 (2010).
\ref [15] N. Jafari and B. Shukirgaliyev, \PL{B853}, 1386093 (2024).
\ref [16] S. Majid, "Foundations of Quantum Group Theory", Cambridge University Press, 1995.
M. Chaichian, A. Demichev, "Introduction to Quantum Groups, World Scientific, 1996.
V.G. Drinfeld, "Quantum groups", Proceedings of the ICM, Rhode Island USA, 1987.
\ref [17] M. Arzano and J. Kowalski-Glikman, \PL{B760}, 69 (2016).
\ref [18] M. Arzano, A. Bevilacqua, J. Kowalski-Glikman, G. Rosati and J. Ungar, \PR{D103}, 106015 (2020).
\ref [19] T. Adach, A. Bevilacqua, J. Kowalski-Glikman, G. Rosati and W. Wi\'slicki, \arx{2507.18336}.
\ref [20] H. Kleinert, "Particles and Quantum fields", World Scientific 2016.
\ref [21] P. Kosi\'nski, J. Lukierski and P. Ma\'slanka, \PR{D62}, 025004 (2000); Cz. J. Phys {\bf 50}, 1283 (2002).
\ref [22] S. Nowak, \hep{0501017}.
\ref [23] M. Daszkiewicz, K Imilkowska, J. Kowalski-Glikman and S. Nowak, \IJMP{A20}, 4925 (2005).
\ref [24] S. Majid and H. Ruegg, \PL{B334}, 348 (1994).
J. Lukierski, H. Ruegg and W.J. Zakrzewski, \AoP{243}, 90 (1995).
\ref [25] D. Kova\v cevi\' c, S. Meljanac, A. Samsarov and Z. \v Skoda, \IJMP{A30}, 1550019 (2015).

\endref
\end